\documentclass{ws-nmnc}
\newcommand{\bhline}[1]{\noalign{\hrule height #1}}

\begin{document}

\markboth{K. Katahira and Y. Chen}
{The Recovery of Hurst Exponent in Speculation Game}

\catchline{}{}{}{}{}

\title{AN EXTENDED SPECULATION GAME FOR THE RECOVERY OF HURST EXPONENT OF FINANCIAL TIME SERIES}

\author{KEI KATAHIRA\footnote{Research Fellow of Japan Society for the Promotion of Science}}

\address{Graduate School of Frontier Sciences, The University of Tokyo,\\ 5-1-5 Kashiwanoha, Kashiwa-shi, Chiba-ken 277-8563, Japan \\
k.katahira@scslab.k.u-tokyo.ac.jp}

\author{YU CHEN}

\address{Graduate School of Frontier Sciences, The University of Tokyo,\\ 5-1-5 Kashiwanoha, Kashiwa-shi, Chiba-ken 277-8563, Japan \\
chen@edu.k.u-tokyo.ac.jp}

\maketitle

\begin{history}
\received{8 November 2018}
\revised{Day Month Year}
\end{history}

\begin{abstract}
The speculation game is an agent-based toy model to investigate the dynamics of the financial market. Our model has achieved the reproduction of 10 of the well-known stylized facts for financial time series. However, there is also a divergence from the behavior of real market. The market price of the model tends to be anti-persistent to the initial price, resulting in the quite small value of Hurst exponent of price change. To overcome this problem, we extend the speculation game by introducing a perturbative part to the price change with the consideration of other effects besides pure speculative behaviors.

\keywords{Cognitive agent-based model; Round-trip trading; Financial stylized facts.}
\end{abstract}

\section{Introduction}	
The financial time series of asset returns have several qualitative properties collectively called stylized facts. For example, one of the well known stylized facts is called heavy tails, which describes that the probability distribution of price returns has fatter tails than those of Gaussian distribution \cite{mantegna2000}$^{,}$\cite{gopikrishnan2000}. Cont summarized 11 currently well known stylized facts including heavy tails \cite{cont2001}. They are quite nontrivial features which can be observed in different markets and instruments.

To investigate the emerging mechanism of stylized facts, we build a novel simple agent-based model named Speculation Game \cite{katahira2019development}, which has two distinct features comparing with preceding agent-based market models. First, it takes account of {\bf round-trip trades} by extending the decision-making structure of Minority Game \cite{challet1997}. Second, a {\bf mutual projection} between players' realistic and cognitive worlds is implemented explicitly. As a result of these novelties, Speculation Game succeeds in reproducing 10 out of the 11 well known stylized facts (see Table \ref{tab1}). 

\begin{table}[htbp]
\caption{The reproducibility of stylized facts in Speculation Game. Symbol ``$+$'' shows that the model successfully recovers the property, and symbol ``$-$'' means that it does not. }
\label{tab1}
\begin{center}
\begin{tabular}{l|c} \bhline{1.1pt}
Stylized fact & Reproducibility\\ \bhline{1.1pt}
Volatility clustering & $+$ \\
Intermittency & $+$ \\
Heavy tails & $+$ \\
Absence of autocorrelation in returns & $+$ \\
Slow decay of autocorrelation in volatilities & $+$ \\
Volume/volatility correlation & $+$ \\
Aggregational Gaussianity & $+$ \\
Conditional heavy tails & $+$ \\
Asymmetry in time scales & $+$ \\
Leverage effect & $+$ \\
Gain/loss asymmetry & $-$ \\ \bhline{1.1pt}
\end{tabular}  
\end{center}
\end{table}

\begin{figure}[htbp]
  \begin{minipage}{0.48\hsize}
  \begin{center}
   \includegraphics[width=62mm]{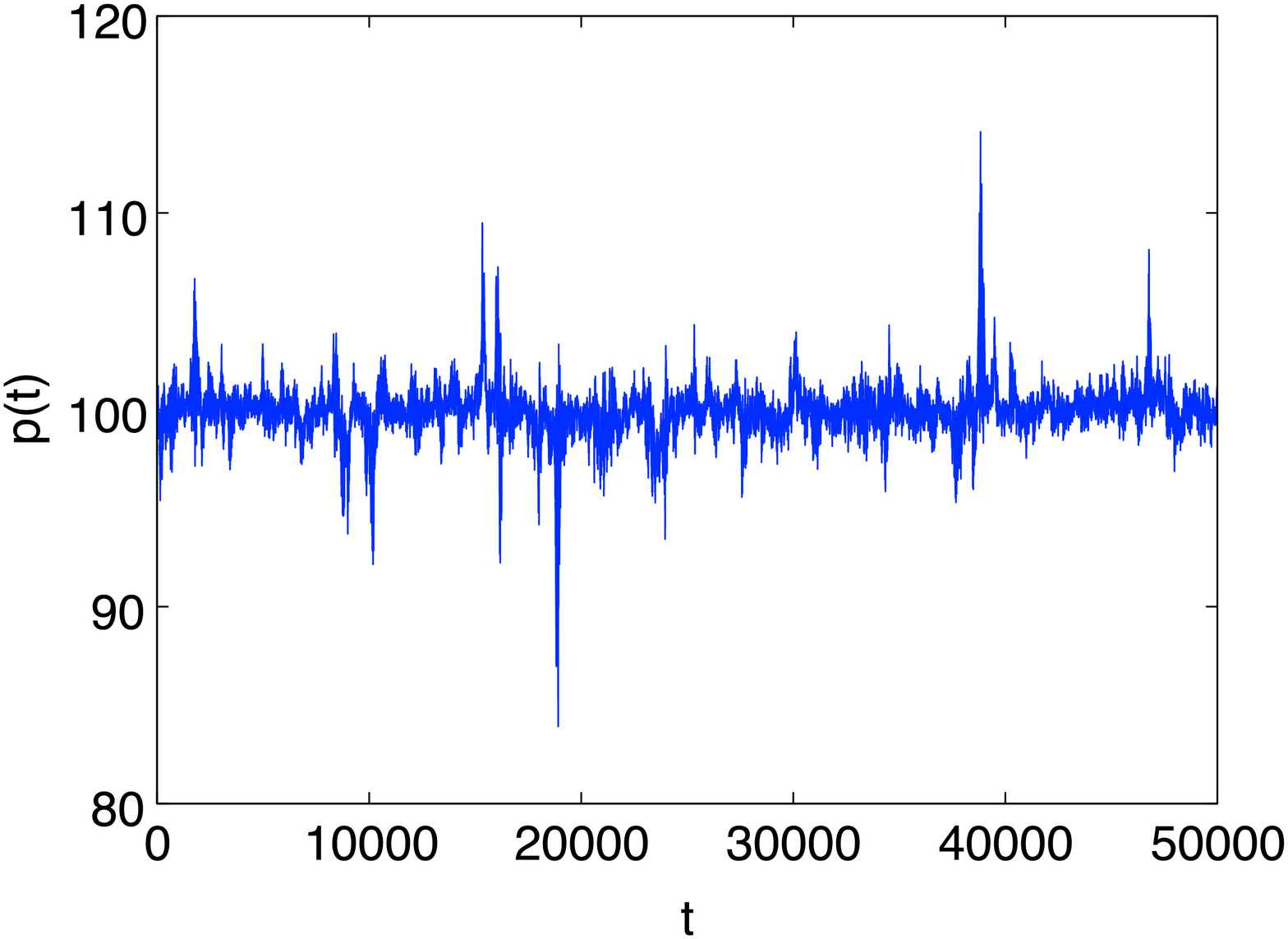}
  \end{center}
  \caption{The tendency of $p(t)$ to stick around the initial price ($=100$) for whole time periods.}
  \label{fig1}
 \end{minipage}
  \hspace{0.02\hsize}
 \begin{minipage}{0.48\hsize}
  \begin{center}
   \includegraphics[width=62mm]{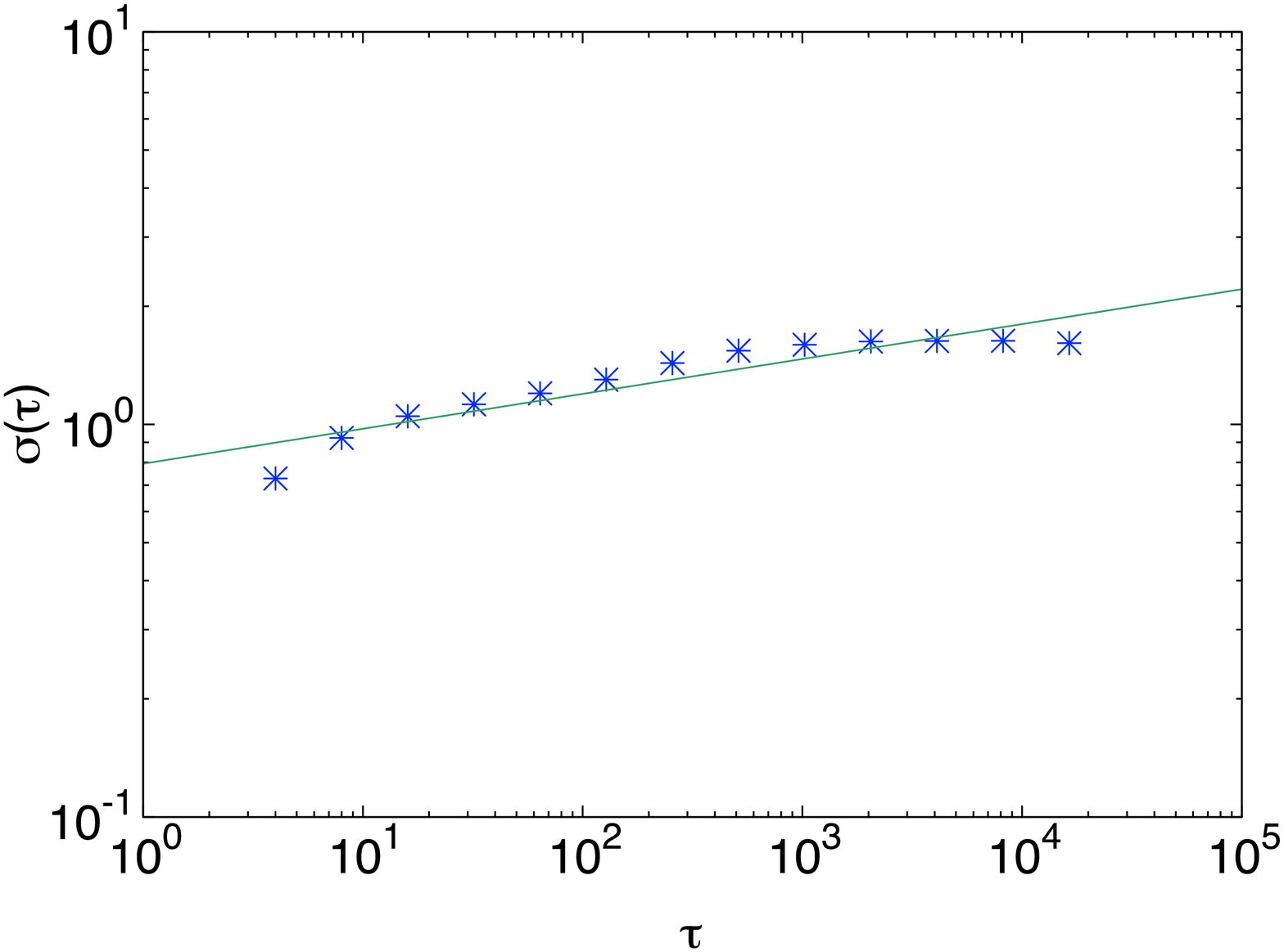}
  \end{center}
  \caption{A poor regression of 100-trial averaged $\sigma(\tau)$ on $\tau$ for 50,000 time steps.}
  \label{fig2}
 \end{minipage}
\end{figure}

However, there is still a divergence from the behavior of real market. The market price $p(t)$ of Speculation Game tends to stick around the initial price as Fig. \ref{fig1} displays (the parameters are fixed as $N=1000$, $M=5$, $S=2$, $B=9$, $C=3$ in this study), which causes an anti-persistent price change with very small Hurst exponents. As a power exponent of regression line of standard deviation $\sigma(\tau)$ (or several other forms) on time scale $\tau$, Hurst exponent $H$ describes the long-term trend in the price time series. Standard deviations of price changes on different time scales can be calculated as follows,
\begin{equation}
\sigma(\tau)=\sqrt{\langle(p(t+\tau)-p(t))^2\rangle - \langle p(t+\tau)-p(t)\rangle^2}.
\label{eq1}
\end{equation}
When $H=0.5$, a price time series is a random walk. When $H>0.5$, it shows a persistent trend while it has a mean-reverting feature when $H<0.5$. Hurst exponent in Speculation Game is found as $H\simeq 0.089$, which indicates its price time series has a too strong mean-reverting characteristic. Note that the fitting of regression is also not very good as $R^2 \simeq 0.86$ in Fig. \ref{fig2}.

Objective of this study is to overcome the price sticking problem and recover Hurst exponent to the normal level like ones in the real financial markets. Particularly, some perturbation is added to price change as the consideration of other effects besides pure speculative behaviors.

\section{Speculation Game and its Extension}

Speculation Game is a repeated game in which $N$ players in a gamified market, with an initial market wealth and randomly given $S$ strategies, compete with each other to increase wealth through round-trip trades. The game proceeds with updates alternating between players’ realistic and cognitive worlds.

At discrete time $t$, player $i$ using her best strategy $j^*$ ($\in S$) takes an action $a_i^{j^*}(t)$ from the set: buy ($=1$), sell ($=-1$), or hold (idle) ($=0$). When she sum it’s an order with strategy $j^*$, the quantity of order $q_i(t)$ is decided both with her market wealth $w_i(t)$ and the board lot amount $B$, the latter of which describes the ease of placing orders with multiple quantities, as follows,
\begin{equation}
q_i(t) = \lfloor \frac{w_i(t)}{B} \rfloor,
\label{eq2}
\end{equation}
where $\lfloor\cdots\rfloor$ stands for a flooring operator. Note that the closing quantity $q_i(t)$ of a round-trip trade is required to be equal to the opening one $q_i(t_0)$. Also, the player's initial market wealth $w_i(0)$ is decided with a uniformly distributed random number $U[0,100)$ to enable her to order one unit at least
\begin{equation}
w_i(0) = \lfloor B+U[0,100) \rfloor.
\label{eq3}
\end{equation}
If $w_i(t)<B$ as a result of round-trip trade, the player will be forced to leave the market and substituted by a new player.

Following the order imbalance equation defined by Cont and Bouchaud \cite{cont2000}, the market price change $\Delta p$ is calculated as follows by letting the initial market price as $p(0)=100$,
\begin{equation}
\Delta p = p(t) - p(t-1) = \frac{1}{N}\sum_{i=1}^{N}a_i^{j^*}(t)q_i(t).
\label{eq4}
\end{equation}
Then, the quantized price movement $h(t)$ (the last digit in the history $H(t)$) is decided by the magnitude correlation between $\Delta p$ and the cognitive threshold $C$, the latter of which can be explained as a threshold value for the players to recognize a big price move: 
\begin{equation}
h(t) = \begin{cases}
2\ ({\rm largely\ up}) & {\rm if}\ \Delta p>C,\\
1\ ({\rm up}) & {\rm if}\ C \geq \Delta p>0,\\
0\ ({\rm stay}) & {\rm if}\ \Delta p=0,\\
-1\ ({\rm down}) & {\rm if}\ -C \leq \Delta p<0,\\ 
-2\ ({\rm largely\ down}) & {\rm if}\ \Delta p<-C.
\end{cases}
\label{eq5}
\end{equation}
To find an action of the best strategy $a_i^{j^*}(t)$, the player with memory $M$ at first takes the reference of the last $M$ digits of history $H(t)$. Next, she looks up strategy $j^*$ to obtain a recommended action corresponding to the historical pattern (see Table \ref{tab2}). However, when the recommendation is the same as the opening order $a_i^{j^*}(t_0)$, the player will hold the position. 

\begin{table}[htbp]
\caption{Example of a strategy for $M=3$. Each strategy table holds recommended actions corresponding to whole historical patterns.}
\label{tab2}
\begin{center}
\begin{tabular}{rrr|rrrrr} \bhline{1.1pt}
\multicolumn{3}{c|}{History} &  \multicolumn{5}{c}{Recommended action}\\ \bhline{1.1pt}
$-2$ & $-2$ & $-2$ & & & & $1$ &\\
$-2$ & $-2$ & $-1$ & & & & $0$ &\\
$-2$ & $-2$ & $0$ & & & & $0$ &\\
$-2$ & $-2$ & $1$ & & & & $-1$ &\\
$-2$ & $-2$ & $2$ & & & &  $1$ &\\
$-2$ & $-1$ & $-2$ & & & &  $0$ &\\
&  \vdots \,  &  & & & & \vdots \, & \\
$2$ & $2$ & $2$ & & & & $-1$ & \\ \bhline{1.1pt}
\end{tabular}  
\end{center}
\end{table}

To determine the best strategy $j^*$, all strategies are evaluated through virtual round-trip trades in the background similarly. Hence, performance of the strategies are assessed in the cognitive world, in terms of the accumulated strategy gains $G_i^j(t)$ ($j \in S$) calculated with cognitive price $P(t)$ corresponding to the rough information of $\Delta p$ in $H(t)$. Letting $P(0) = 0$, $P(t)$ is updated as 
\begin{equation}
P(t) = P(t-1)+h(t).
\label{eq6}
\end{equation}
The gain of strategy $j$ in a round-trip trade $\Delta G_i^j(t)$ reads as
\begin{equation}
\Delta G_i^j(t) = a_i^j(t_0)(P(t) - P(t_0)).
\label{eq7}
\end{equation}
Thus, the accumulated strategy gain $G_i^j(t)$ is calculated by:
\begin{equation}
G_i^j(t) = G_i^j(t_0) + \Delta G_i^j(t).
\label{eq8}
\end{equation}
Whenever the accumulated gain of the strategy in use $G_i^{j^*}(t)$ is updated, all $G_i^j(t)$ will be reviewed to renew the strategy with the best performance. If the best strategy happens to be one of the unused strategies with a virtual trade ongoing, the virtual position will be closed immediately before the player switches the strategy at the next time step. Note that the evaluating system is developed by considering that the investing strategies should be evaluated through by round-trip trades, as Katahira and Akiyama pointed out \cite{katahira2017}.

Since the self-financing assumption is not required in Speculation Game, when a round-trip trade is closed in the realistic world, the player's market wealth $w_i(t)$ is updated with an investment adjustment $\Delta  w_i(t)$, which is the conversion of strategy gain  in the cognitive world with $q_i(t)$ taken into consideration,
\begin{equation}
w_i(t) = w_i(t_0) + \Delta  w_i(t) = w_i(t_0) + f(\Delta G_i^{j^*}(t)q_i(t_0)),
\label{eq9}
\end{equation}
where $f$ can be an arbitrary function. In the game, $\Delta  w_i(t) = \Delta G_i^{j^*}(t)q_i(t_0)$ is used for the simplicity. 

As an extension to the original model, a random perturbation $U[-Pb,Pb)$ is further added to $\Delta p$ as
\begin{equation}
\Delta p = \frac{1}{N}\sum_{i=1}^{N}a_i^{j^*}(t)q_i(t)+U[-Pb,Pb),
\label{eq10}
\end{equation}
where the first term of Eq. (\ref{eq10}) could be considered as the effect given by the pure speculative players while the second one as other effects by other types of traders such as long term value traders, hedgers, and so on. Due to this extension, $\Delta p$ never takes zero so that $H(t)$ become a quaternary time series instead of a quinary one.

\section{Result and Discussion}
\begin{figure}[htbp]
 \begin{minipage}{0.48\hsize}
  \begin{center}
   \includegraphics[width=62mm]{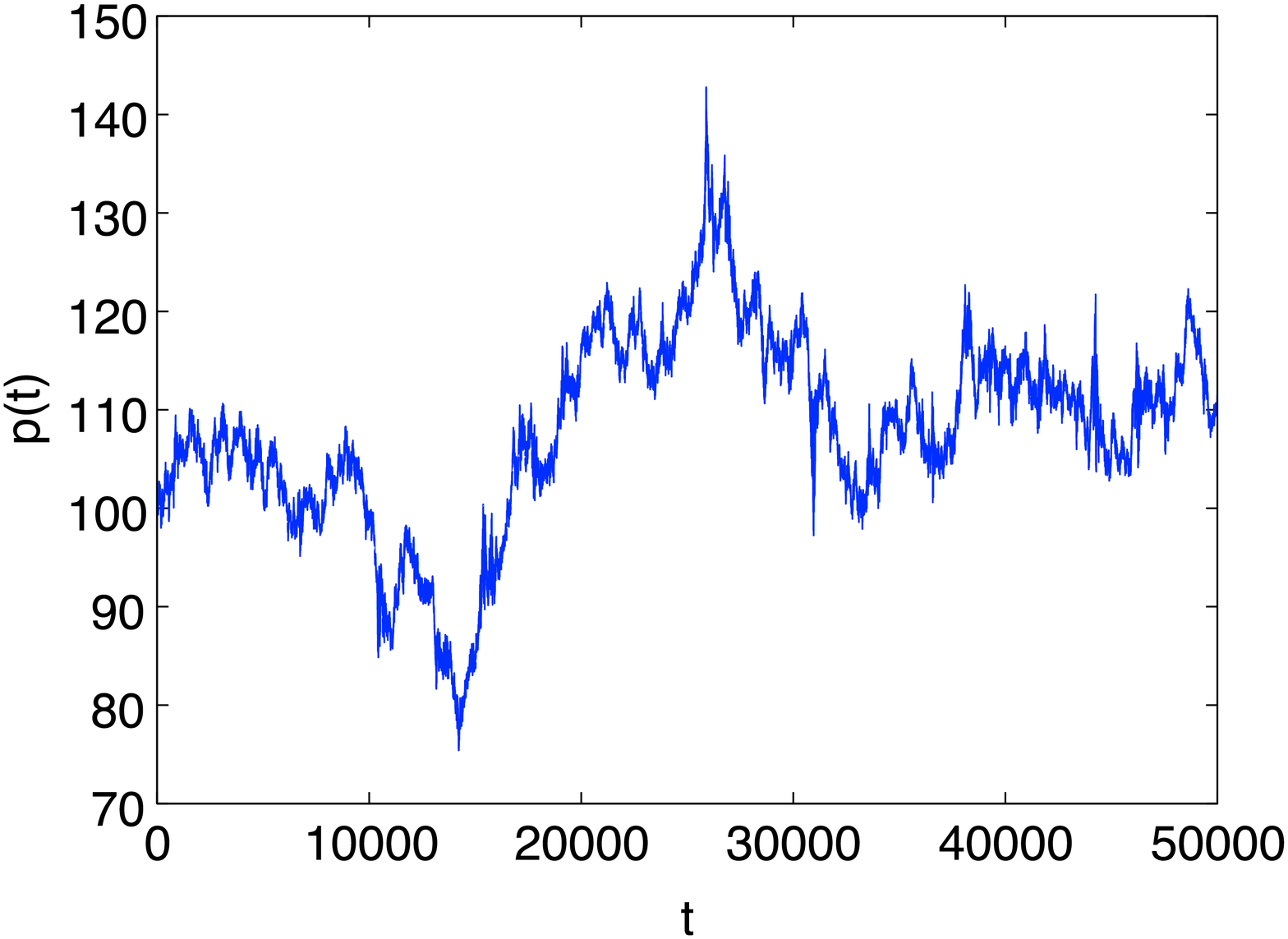}
  \end{center}
  \caption{The more realistic movement of $p(t)$ with long-term fluctuations when $Pb=0.25$.}
  \label{fig3}
 \end{minipage}
  \hspace{0.02\hsize}
 \begin{minipage}{0.48\hsize}
  \begin{center}
   \includegraphics[width=62mm]{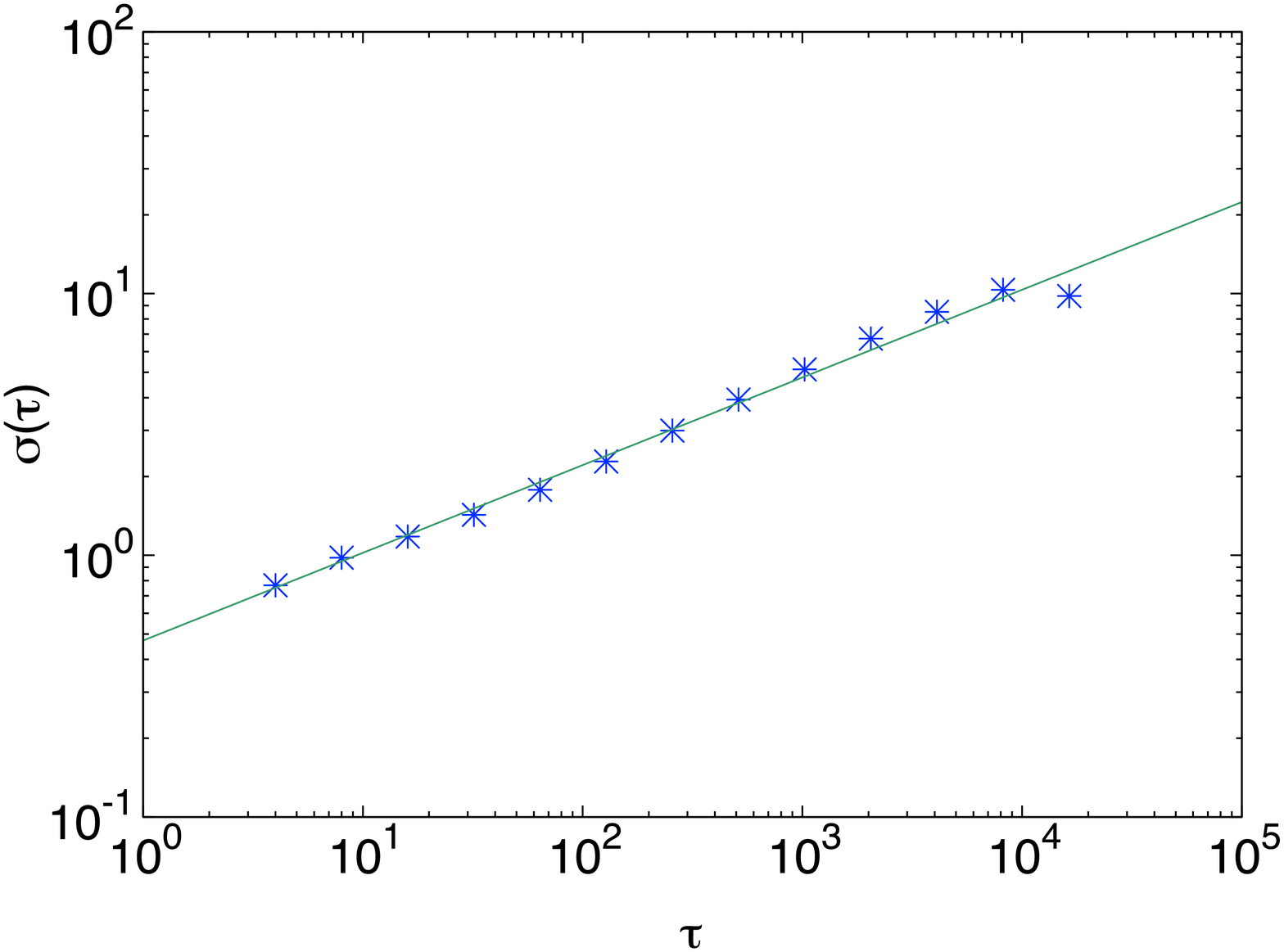}
  \end{center}
  \caption{The recovery of Hurst exponent and the better regression when $Pb=0.25$.}
  \label{fig4}
 \end{minipage}
\end{figure}

By adding the random perturbation, the dynamics of market price in Speculation Game becomes more realistic. As Fig. \ref{fig3} displays, $p(t)$ can escape from the initial price range and shows long-term fluctuating structures when $Pb=0.25$. Accordingly, Hurst exponent also increases to the normal level ($H \approx 0.3 \sim 0.7$) as $H\simeq 0.33$ with a better fitting ($R^2 \simeq 0.99$) as the regression line shown in Fig. \ref{fig4}. It can also be inferred from these results that the long-term fluctuations and trends originated from those non-speculative behaviors. 

\begin{figure}[htbp]
 \begin{minipage}{0.48\hsize}
  \begin{center}
   \includegraphics[width=62mm]{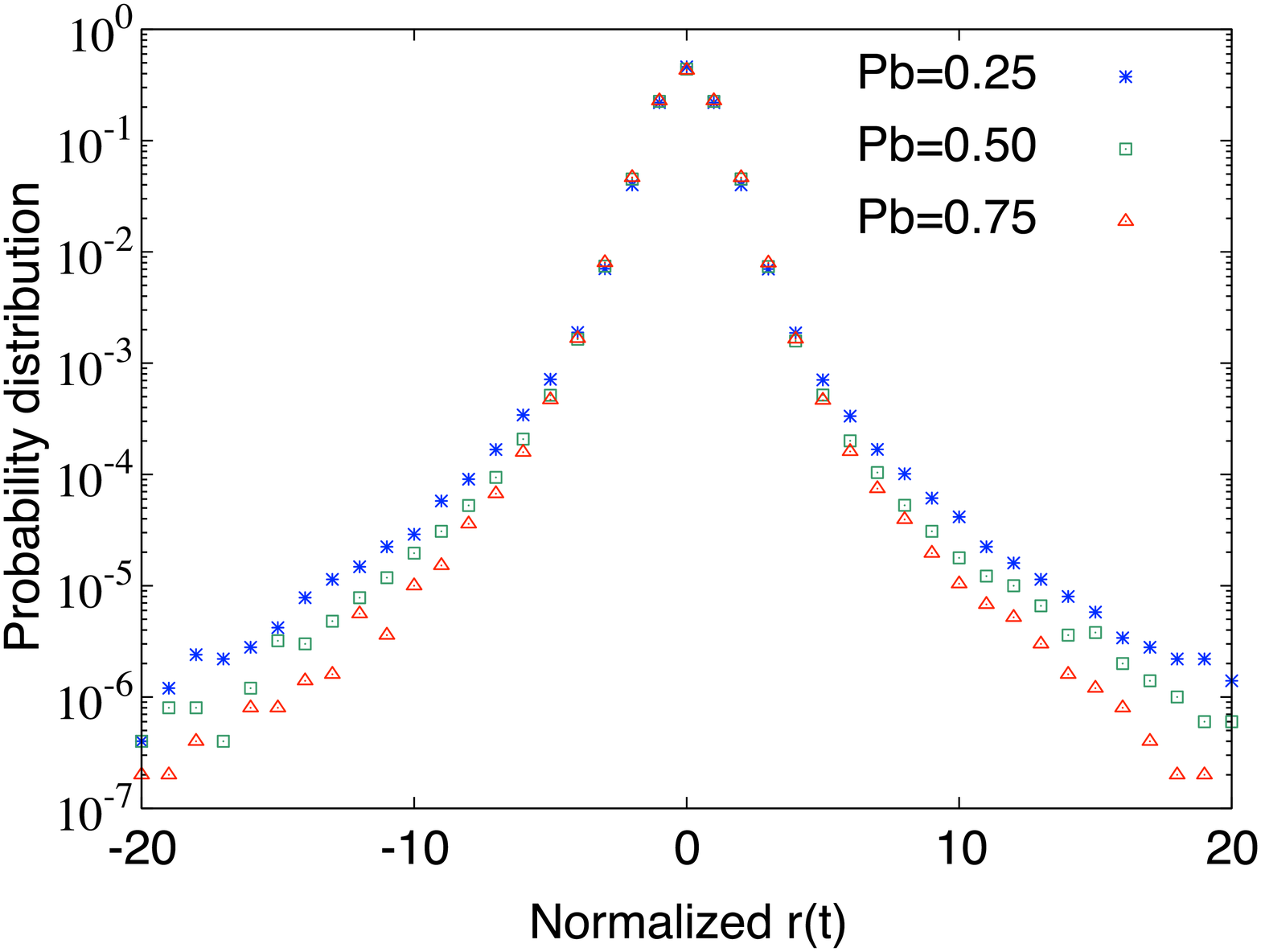}
  \end{center}
  \caption{The thinner tails of return distributions by increasing $Pb$ as 0.25, 0.50, and 0.75.}
  \label{fig5}
 \end{minipage}
  \hspace{0.02\hsize}
 \begin{minipage}{0.48\hsize}
  \begin{center}
   \includegraphics[width=62mm]{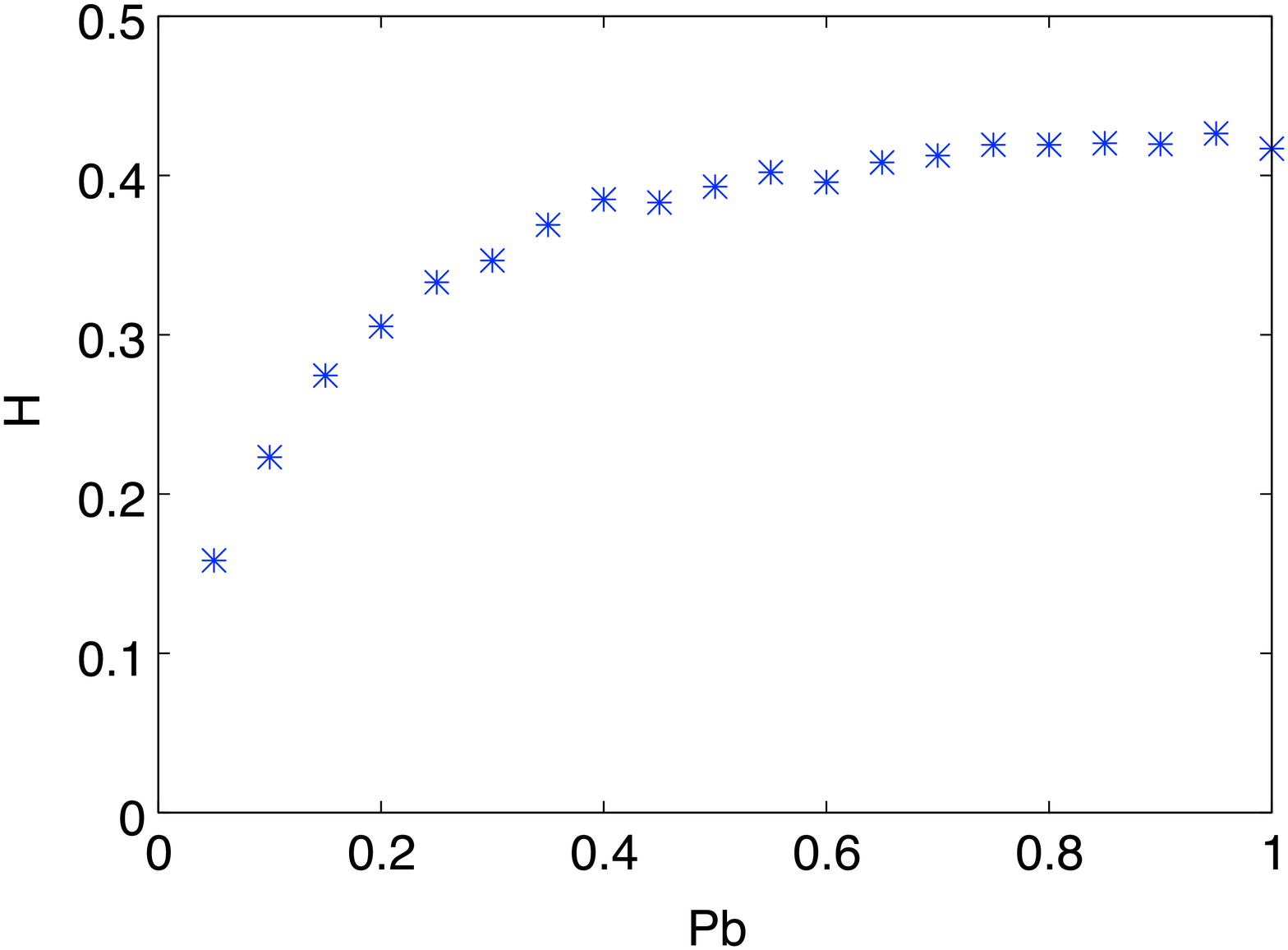}
  \end{center}
  \caption{The decay in the increment of Hurst exponent as $Pb$ increases.}
  \label{fig6}
 \end{minipage}
\end{figure}

On the other hand, the further increment of perturbation would weaken the stylized properties as well. For instance, the probability distributions of price returns have less heavier tails as more perturbation is added, which is displayed in Fig. \ref{fig5}. Since the similar propensity is confirmed for many of other reproduced stylized features in Speculation Game, speculative behaviors can be considered as the main source for the emergence of stylized facts. Note that adding moderate price perturbation ($Pb\simeq 0.2\sim 0.3$) is favorable in the viewpoint of efficiency to enhance Hurst exponent because excess $Pb$ contribute less to the increment of $H$ as Fig. \ref{fig6} shows. Also, it is reasonable that the perturbation can not make $H>0.5$ because of pure random walk nature.

\section{Conclusion and Future Work}
To conclude, adding moderate perturbation to $\Delta p$ can solve the price sticking problem and recover Hurst exponent for our Speculation Game. Simulation results also indicate that the non-speculative behaviors generate the long-term fluctuations and trends while speculative ones contribute to the emergence of well-known financial stylized facts.

For future research, decision-making structures for fundamentalists and news traders need to be developed for the generation of a long-term trend. Adding perturbation is the simplest way to include the effects brought by those non-pure speculative behaviors, though these effects should be further verified in a more concrete way as well.

\section*{Acknowledgments}
This work was supported by JSPS KAKENHI grant number JP17J09156.

\end{document}